\documentclass{ieeeaccess}
\usepackage{epstopdf}
\usepackage{cite}
\usepackage{amsmath,amssymb,amsfonts}
\usepackage{algorithmic}
\usepackage[lined, linesnumbered, ruled]{algorithm2e}
\usepackage{graphicx}
\usepackage{textcomp}
\usepackage{caption}
\usepackage{soul}
\usepackage[none]{hyphenat}
\def\BibTeX{{\rm B\kern-.05em{\sc i\kern-.025em b}\kern-.08em
    T\kern-.1667em\lower.7ex\hbox{E}\kern-.125emX}}
\begin{document}
\history{Received October 7, 2018, accepted November 7, 2018, date of publication November 19, 2018, date of current version December 18, 2018.}
\doi{10.1109/ACCESS.2018.2882018}

\title{CMIP: Clone Mobile-agent Itinerary Planning Approach for Enhancing Event-to-Sink Throughput in Wireless Sensor Networks}
\author{\uppercase{Huthiafa Q Qadori}\authorrefmark{1}, \uppercase{Zuriati Ahmad Zukarnain}\authorrefmark{1},\IEEEmembership{Member, IEEE}, \uppercase{Zurina Mohd Hanapi}\authorrefmark{1}, \IEEEmembership{Member, IEEE}, \uppercase{Shamala Subramaniam}\authorrefmark{1,2}\IEEEmembership{Member, IEEE}, and\uppercase{ MOHAMED A. ALRSHAH}\authorrefmark{1}\IEEEmembership{Senior Member, IEEE}}
\address[1]{Department Communication Technology \& Network, Faculty of Computer Science \& Information Technology, Universiti Putra Malaysia, Serdang 43400, Malaysia}
\address[2]{Sports Academy, Universiti Putra Malaysia, Serdang 43400 UPM, Malaysia}

\tfootnote{This work is supported by Universiti Putra Malaysia, the Ministry of Higher Education-Iraq (University of Anbar)}

\markboth
{Huthiafa Q Qadori \headeretal: Preparation of Papers for IEEE Access}
{Huthiafa Q Qadori \headeretal: Preparation of Papers for IEEE Access}

\corresp{Corresponding authors: Huthiafa Q Qadori (Huthiafaqadori@gmail.com) and Zuriati Ahmad Zukarnain (zuriati@upm.edu.my)}

\begin{abstract}
In order to mitigate the problem of data congestion, increased latency, and high energy consumption in Wireless Sensor Networks (WSNs), Mobile Agent (MA) has been proven to be a viable alternative to the traditional client-server data gathering model. MA has the ability to migrate among network nodes based on an assigned itinerary, which can be formed via Single Itinerary Planning (SIP) or Multiple Itinerary Planning (MIP). MIP-based data gathering approach solves problems associated with SIP in terms of task duration, energy consumption, and reliability. However, the majority of existing MIP approaches focus only on reducing energy consumption and task duration, while the Event-to-sink throughput has not been considered. In this paper, a Clone Mobile-agent Itinerary Planning approach (CMIP) is proposed to reduce task duration while improving the Event-to-sink throughput in real-time applications, especially when the MA is assigned to visit a large number of source nodes. Simulation results show that the CMIP approach outperforms both Central Location-based MIP (CL-MIP) and Greatest Information in Greatest Memory-based MIP (GIGM-MIP) in terms of reducing task duration by about 56\% and 16\%, respectively. Furthermore, CMIP improves the Event-to-sink throughput by about 93\% and 22\% as compared to both CL-MIP and GIGM-MIP approaches, respectively.
\end{abstract}

\begin{keywords}
Data gathering, Clone, Mobile Agent, Static Itinerary, Dynamic Itinerary, MIP
\end{keywords}

\titlepgskip=-15pt

\maketitle

\section{Introduction}
\label{sec:introduction}
The merging of low-cost wireless communication, computation, and sensing have spawned a new generation of small and cheap intelligent devices. The deployment of tens to thousands of these tiny devices in self-organizing networks has established a new network type known as Wireless Sensor Networks (WSNs). A WSN can be generally described as the deployment of a vast number of tiny sensor nodes that connect to each other via wireless communication. The advantages of these sensor nodes; such as easy installation, low cost, small size and low power consumption; make this type of networks very useful for many applications in different fields of life such as agriculture, industrial automation, transportation, health care, and military \cite{puccinelli2005wireless}.

The main objective of WSN deployment is to gather environmental data and deliver it to end users \cite{al2004routing}. Typically, in a WSN, sensor nodes are deployed in a field of interest where every sensor node is able to perform several tasks such as sensing, surveillance, environmental monitoring, processing and storing the sensed data. The sensed data at any sensor node needs to be transmitted to a collection point in the network, called a processing unit or sink, for further analysis. The transmission of data is done via single or multi-hop wireless communication. This type of transmission is based on the conventional client-server model, where every involved sensor node transmits data packets to the sink.

The WSNs have a very limited link bandwidth compared to wired networks, a set of connected wireless links need to be established from those nodes to the sink in order to forward data packets. Consequently, this could increase network data traffic and, as a result, it could increase the consumption of network resources. As a solution to this overwhelming data traffic, \cite{qi2003} has proposed an MA-model for data gathering in WSNs. In this model, a software component, called MA, is forwarded throughout the network and to visit source nodes one by one following an itinerary to perform data gathering process locally at each source node. Upon completion of data gathering, the MA brings back the aggregated data to the sink. Consequently, this data combination decreases the number of packets transmitted to the sink, which results in a decrease in communication costs, bandwidth utilization, network congestion, and energy consumption.

In WSNs, MA-based data gathering employs two schemes of the itinerary; Single Itinerary Planning (SIP) and Multi Itinerary Planning (MIP). SIP utilizes single MA to roam networks for data gathering \cite{qi2001,chen2007A,wu2004,chen2009}, while MIP utilizes multiple MAs that work in parallel to perform data gathering \cite{cai2011,bendjima2012A}. In large-scale networks, SIP introduces many drawbacks \cite{qadori2017} such as long delays, increase in MA packet size, low reliability, and high probability of MA packet loss, due to the migrating to a large number of source nodes.

To overcome these drawbacks, MIP \cite{cai2011,bendjima2012A} was proposed, in which several MAs are distributed into the network to visit groups or partitions of source nodes concurrently, where each MA is assigned to only one group of source nodes. This process reduces the MA packet size, which further leads to a lower energy consumption compared to SIP schemes. Moreover, the task duration in MIP is minimized, due to the distribution of concurrent aggregation tasks among multiple MAs.

As known, the MA itinerary is the source-visiting sequence that the MA needs to follow during its migration, where the MA migration itinerary planning is still a challenging issue in MA-based data gathering. In spite of MIP advantages, it worth noting that finding an optimal itinerary for the MA is a Nondeterministic Polynomial (NP-complete or NP-hard) problem \cite{chen2007, chen2009}. A sub-optimal MA itinerary may lead to a highly inefficient overall network performance.

In MA-based data gathering, the determination of MA's itinerary can be classified as static, dynamic or hybrid itinerary \cite{chen2007,mpitziopoulos2009}. In static itinerary, the sequence of source nodes that will be visited by MA is calculated at the sink before the MA starts its migration, while in the dynamic itinerary, the sequence of source nodes that will be visited by MA is determined on the fly at each source node. In hybrid itinerary, the source nodes to be visited are selected at the sink, but the visiting sequence is computed on the fly by the MA. However, dynamic or hybrid itineraries consume valuable node energy resources and employ larger MA. This is because the MA needs to carry the next hop computation code to be executed at each node during the migration process. On the other hand, static itinerary consumes less energy compared to dynamic or hybrid itinerary since the MA carries only a pre-determined itinerary that has been calculated at the sink. 

The applications of WSNs are designed to accomplish specific objectives or desired tasks. These objectives or tasks are defined and optimized according to the application user. The main goal of WSNs is to deliver the sensed data to the relevant processing unit (sink) for further analysis and decision making, where the collected data must reflect the current state of the targeted environment. In most cases, the collected data is valid only for a limited period of time due to the constant changes in the monitored environment. Thus, the importance of these data changes from accurate to less accurate until it becomes totally inaccurate to reflect the current state of the environment as time progresses \cite{diallo2012real}. In real-time WSNs' applications, it is very crucial to deliver the collected data to the sink without delay to make timely actions or decisions. Moreover, collecting as much data as possible is also very important to the WSN application user to allow him or her to improve the quality of evaluation and analysis to make precise decisions \cite{pourroostaei2014wireless}. 

Indeed, MIP approaches focus only on reducing energy consumption and task duration, while neglecting the amount of delivered data to the sink within the time period. However, collecting as much data as possible requires the distributed MAs to visit a larger number of source nodes, which definitely will increase the delay. To solve this issue, a MIP approach that is able to minimize task duration and maximize collected data is highly needed to support the real-time applications. However, this solution will be at the expense of energy efficiency. 

This paper proposes a novel MIP-based data gathering approach in WSNs, namely CMIP, which aims to reduce task duration of data gathering process while maximizing the volume of collected data. The remainder of the paper is organized as follows: Section II explains the related works, and Section III presents the proposed CMIP approach, while Section IV presents the simulation setup. Further, the performance evaluation and experimental results are explained in Section V. Finally, Section VI concludes the work and presents the future directions for the potential improvements.

\section{Related works}
In the last few years, several MIP-based data gathering approaches have been proposed. The main objective of those approaches was to reduce the energy consumption and task duration. This section reviews several MIP-based approaches proposed for data gathering, discusses their basic concepts, and highlights their advantages and shortcomings.

In \cite{mpitziopoulos2007}, a Near-Optimal Itinerary Design (NOID) algorithm was proposed to address the problem associated with calculating the number of near-optimal routes for MAs. NOID algorithm adapts the Esau-Williams heuristic method \cite{esau1966} that were designed for the Constrained Minimum Spanning Tree (CMST) problem in network designing. NOID algorithm iteratively groups sensor nodes in the network to separate sub-trees that are connected progressively to the Processing Element (PE) or sink. Finally, each sub-tree is assigned to an individual MA, where the number of needed MAs is equal to the number of trees in the relevant network.

Later, the Tree-Based Itinerary Design (TBID) algorithm \cite{konstantopoulos2010} was proposed, which outperformed NOID algorithm in terms of low-cost itineraries. TBID not only determines the optimal number of MAs but also creates low-cost itineraries for each individual MA, by partitioning the area around the PE/sink into concentric zones. The number of nodes lying within the radius of the first zone of PE/sink represents the starting points of itineraries for the needed MAs. The radius of the first zone can be calculated as ${a r}_{max}$, where $r_{max}$ is the maximum transmission range of any sensor node and the input parameter $a \in (0, 1]$. The subsequent zones width is calculated as $r_{max}/2$, such that each sensor node in any zone can only communicate with sensor nodes belonging to the previous, current, and next outer zones. In each iteration, MAs itineraries start from inner zones and proceed to outer zones connecting the pair of adjacent sensor nodes $u$ and $v$, where $u$ is linked to an itinerary and $v$ is not yet linked. This minimizes the edge Potential Cost (PC), which is equal to the cost of an itinerary derived in the case of incorporating the edge in the itinerary. This ensures low-cost itineraries to be created for each individual MA. The end output of TBID is $k$ trees, where each tree is rooted at the nodes lying within the radius of the first zone of PE/sink. The itinerary of each MA in these trees is derived by a post-order traversal.

\cite{gavalasEnergy} introduced a novel algorithm for the energy-efficient itinerary planning of MAs. This algorithm adopts a meta-heuristic method called Iterated Local Search (ILS) to derive the hop sequence of multiple traveling MAs over the deployed source nodes. Like other tree-based MIP algorithms, ILS determines the number of itineraries (MAs) by considering a circular zone around the sink. The nodes that are lying in the sink zone will be the starting points of each MA itinerary. However, as opposed to other tree-based MIP algorithms, ILS iteratively examines the energy cost of potential attachment of each candidate node $u$ (among those that remain unconnected) with any pair of already attached subsequent nodes. Such that the minimum itinerary cost among the examined candidate nodes will be added to the current MA's itinerary. As a result, the ILS approach builds low energy cost of MAs' itineraries.

Although NOID, TBID, and ILS perform better than SIP approaches, the MA in these tree-based schemes consumes twice as much energy due to the reverse routes that the MA take, especially when there are many branches. This results in an increase in the energy consumption and task duration of MA's migration.

The central location-based MIP (CL-MIP) is another algorithm proposed by \cite{chen2009A}, where the determination of the optimal number of MAs can be divided into four parts; (1) Visiting Central Location (VCL) selection algorithm, (2) Source grouping algorithm,  (3) Source-visiting order determination SIP-based  algorithm, (4) Iterative algorithm; to ensure that all source nodes have been assigned to their MAs. CL-MIP groups all source nodes according to the node density (gravity algorithm). The basic idea of VCL algorithm is to distribute each source node's impact factor to other source nodes. Let $n$ represent the source node number; then each source node will receive ($n$- 1) impact factors from other source nodes, and one from itself. At each iteration, the location of the source node with the highest accumulated impact factor will be selected as a VCL. Then, all source nodes within the radius of VCL are grouped in a cluster and assigned to an MA. The above process repeats until all remaining source nodes in the network are assigned to their MAs, then, the itinerary for each MA can be planned by a SIP algorithm. In CL-MIP, the itinerary for each distributed MA is determined by local closest first (LCF) \cite{qi2001}. In LCF, the MA looks for the next hop node with the shortest distance from the current location. However, CL-MIP algorithm assumes that the relevant source nodes are geographically distributed in several clusters, which limits the use of this algorithm in a broad range of applications.

In \cite{rais2018determination}, the author proposed an Optimal Multi-agents Itinerary Planning (OMIP) algorithm similar to CL-MIP, where the source nodes are grouped into clusters. OMIP adopts Efficient Clustering Routing Protocol (ECRP) to partition the network into $v$ clusters, select a medoid node $MNi$ in each cluster, and then minimizes the average distance between $MNi$ and other nodes in the cluster. The $MNi$ in OMIP functions as the cluster-head as well as the starting and arrival node of the MA. In OMIP, the sink dispatches an MA to every cluster in the network. The MA starts its data gathering process from $MNi$, roams the source nodes within the cluster and returns back to the $MNi$. After the MA completes data gathering, it travels back to the sink with the aggregated data. OMIP approach suffers from the same drawbacks of CL-MIP, where the clustering method highly depends on the distribution of source nodes. OMIP generates clusters with few source nodes to be visited by the MA. As a result, MAs' itineraries becomes imbalanced which directly affects the network performance in terms of energy consumption and task duration.

Multi-mobile Agent itinerary planning-based Energy and Fault aware (MAEF) was proposed by \cite{el2018multi}. Similar to (OMIP), MAEF also groups the source nodes into clusters and each cluster has a Cluster Head (CH). It adopts the same idea presented in CL-MIP in which a distribution of density impact factor is used to select the CHs. Also, all source nodes, which lie within the maximum transmission range of each selected CH, are grouped together. MAEF differs from CL-MIP and OMIP, as the MAs' itineraries are planned only among the CHs and determined using the MST. Furthermore, the number of itineraries are equal to the number of nodes located within the sink' transmission range. In MAEF, once the MA reaches the CH for the first time, it notifies the CH's source nodes to send their data to the CH. When the MA reaches the last CH in its itinerary, it starts gathering data from CHs on its way back to the sink.

Although the MAEF algorithm deploys a new MIP strategy, where the distributed MAs visit only CHs, it has two main drawbacks. First, the CHs' transmission range (which used to group the source nodes into clusters) can generate overlap between the clusters which can increase the overhead. Second, each MA needs to traverse its itinerary twice; to notify the CH's source nodes to send data, and to gather data from the CHS. This would result in an increase of energy consumption and task duration.

In most of the aforementioned MIP approaches, the geographic information of the sensor nodes is the main parameter used to determine the optimal number of MAs and their itineraries. In \cite{aloui2015}, the Greatest Information in the Greater Memory-based MIP (GIGM-MIP) algorithm was proposed. The GIGM-MIP algorithm considers geographic information to partition the network and takes into account the data size in each partition to formulate the optimal number of MAs. 


In GIGM-MIP, the data size in each partition plays an important role in determining the number of MAs. After partitioning the network, GIGM-MIP calculates the data size of the source nodes in each partition and starts with an initial number of MA. Then, GIGM-MIP updates this number by assigning source nodes to the MAs that have free payload memory size, such that each partition may have more than one MA. Using this strategy, GIGM-MIP approach balances the data load among distributed MAs, thereby reducing task duration. However, exchanging the assigned source node among the MAs results in long MA itinerary, especially when the location of the assigned source node is far from the current MA's location. Moreover, distributing more than one MAs to one partition means that each distributed MA should carry its processing code (aggregation code). Multiple MAs carrying the same aggregation code within a single partition would result in an increase in the MA's migration hops, which subsequently increases the energy consumption of the network. 

In the above-reviewed MIP approaches, many solutions have been proposed to determine the number of MAs and their itineraries, which is the main challenging issue associated with MIP paradigm. Majority of the proposed MIP approaches are focusing on minimizing the energy consumption and task duration, while the amount of the data delivered to the sink within the time period is omitted. In some WSNs applications (e.g. real-time applications) a MIP approach that is able to minimize the task duration and maximize the data collected at the sink is really needed even if it is not optimizing the energy consumption. The proposed CMIP approach in this paper adopts the cloning concept of the MA to reduce the task duration of data gathering process while maximizing data collection.

\section{CMIP Approach}
This section presents the proposed CMIP, which includes two parts, the MA's cloning mechanism and the mechanism of determining the itinerary used by MA(s) to visit the source nodes. 

\subsection{Cloning Mechanism of MA in CMIP}
As mentioned in \cite{chen2009A,cai2011}, the task duration in MIP is the delay experienced by the last MA, which returns to the sink node. In other words, this MA has the longest delay among the other dispatched MAs. In \cite{chen2007A}, the delay of MA's itinerary to complete its data gathering task ($T_{MA}$) can be expressed as follows:
\begin{equation}
\label{T_P_MA}	
T_{MA} = T_{p} + T_{roam} + T_{back}, 
\end{equation}
where $T_{p}$ is the delay of the MA migration from the sink to the first source node, $T_{roam}$ is the MA migration delay from the first source node to the last source node, and $T_{back}$ is the MA migration delay from the last source node to the sink. The $T_{p}$ can be simplified as below: 
\begin{equation}
\label{Tp}
T_{p} = \Big(\frac{pc}{D_r}+ t_{ctrl}\Big)\times H,
\end{equation}
where $pc$ is the MA's size (processing code plus MA packet header), ${D_r}$ is the data rate at MAC layer, $t_{ctrl}$ is the delay of control messages and $H$ is the number of hops between the sink and the first source node. The $T_{roam}$ can be expressed as follows:
\begin{equation}
\label{Troam}
T_{roam} = \sum_{i=1}^{M}\Big(t+\frac{S_{data}}{D_p}+\frac{S_{ma}^{i}+pc}{D_r}+ t_{ctrl}\Big)\times H,
\end{equation}
where $M$ is the number of source nodes, $t$ is the MA access delay (the time required to mount MA's processing code in the target source node), $S_{data}$ is the size of sensed data at the source node, $D_p$ is the data processing rate, and $S_{ma}^{i}$ is the size of the aggregated data carried by the MA at source node~$i$. The $T_{back}$ can be decomposed into:
\begin{equation}
T_{back} = \Big(\frac{S_{ma}^{M}+pc}{D_r}+ t_{ctrl}\Big)\times H,
\end{equation}
where $S_{ma}^{M}$ is the size of MA at the last source node.  

Based on Equation \ref{Troam}, the delay of MA is highly associated with the number of source nodes and the number of MA's hops. The more source nodes to be visited by MA, the larger MA's hops which consequently implies a higher delay. Additionally, when the MA visits a large number of source nodes, its data payload size becomes large, which can further increase the MA delay.
 
\begin{figure*}[h!]
	\begin{center}
		\includegraphics[width=0.85\linewidth]{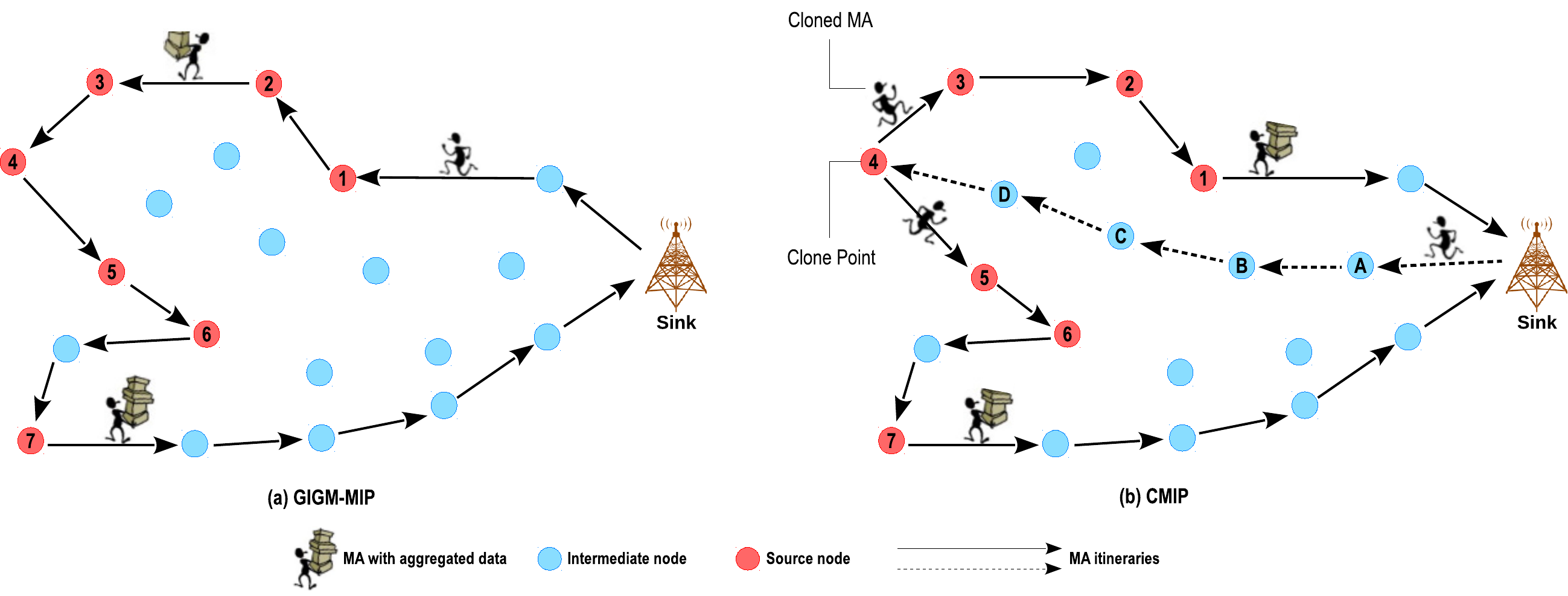}
	\end{center}
	\vspace{-0.5cm}{\textbf{\caption{MA itinerary planing based on: (a) GIGM-MIP; (b) CMIP}
	\label{GIGMvsCMIP}}}
	\addtocontents{lof}{\protect\addvspace{.5cm}}
\end{figure*}
 
The time required by an MA to collect data from a large number of source nodes can be minimized by dispatching two MAs since each MA will be assigned to fewer source nodes. However, this strategy would lead to increased energy consumption due to the large number of hops utilized by the two separated itineraries. Moreover, based on Equation \ref{Tp}, using two separated itineraries would mean each MA is required to carry its own processing code (aggregation code), which further leads to increase energy consumption. This excessive increase in energy consumption would lead to an increase in the probability by which MA could drop its accumulated data. Therefore, there is a need for a solution that distributes the data collection task while reducing the task duration and maximizing the data collection. 

To mitigate the above-mentioned issues, this paper adopt the MA cloning concept \cite{mpitziopoulos2010cbid,javadi2012clone,gupta2014}. In this concept, the MA has the ability to clone itself at a certain point resulting in an instance with the same capacities and capabilities. In the proposed CMIP, one MA is dispatched from the sink, where this MA then clones itself at a certain point (named cloning point). Figure \ref{GIGMvsCMIP} illustrates the MA itinerary planning based on the cloning mechanism in our proposed CMIP as compared to GIGM-MIP approach.

In Figure \ref{GIGMvsCMIP}, there are seven source nodes (the red nodes \textbf{1}, \textbf{2}, \textbf{3}, \textbf{4}, \textbf{5}, \textbf{6}, \textbf{7}) to be visited by the MA. In GIGM-MIP (Figure \ref{GIGMvsCMIP}(a)), the sink has determined one MA to visit the seven source nodes. Using LCF algorithm, the visiting order is such that MA begins from source node \textbf{1}, \textbf{2}, \textbf{3}, \textbf{4}, \textbf{5}, \textbf{6} and finishes at source node \textbf {7}. It is clear that the MA's data payload would have become large after the MA completes data collection from source node \textbf {7}. With this large MA's data payload, the MA needs to migrate 5 hops until it reaches the sink, which increases its delay. On the other hand, Figure \ref{GIGMvsCMIP}(b) shows the MA itinerary planing based on cloning mechanism in CMIP, where the source nodes will be separated into two groups. One group will be assigned to the Main MA (MMA), the MA that has the ability to clone itself, while the second group will be assigned to the Cloned MA (CMA). At the initial stage, the sink determines the Farthest Source Node (FSN), the source node located at the farthest distance from the sink. This FSN can be calculated using Equation (\ref{FSNlocation}), as below:
\begin{equation}
\label{FSNlocation}
FSN=\sqrt{(CSN_x-Sink_x)+(CSN_y-Sink_y)},
\end{equation}
where ($CSN_x$, $CSN_y$) represent the position of a source node points in MA's itinerary, and ($Sink_x$, $Sink_y$) determine the position of the sink.

The FSN is selected to be the cloning point, where the MMA clones itself (for example node number \textbf{4} in Figure \ref{GIGMvsCMIP}(b)). At the FSN, both MMA and CMA will start data collection from the source nodes. The reason behind selecting the FSN to be the cloning point because the operation of data gathering from the farthest source node first towards the sink consumes less energy and time as compared to the one starts from the nearest source node (more details are discussed in the following section).

\subsection{MA's itinerary determination in CMIP}
The delay and energy consumption of an MA itinerary depends on the efficiency of source nodes visiting order. In GIGM-MIP, this visiting order is determined by the LCF algorithm, which selects the nearest source node to the current MA location as the next MA's hop. However, selecting the closest source node as the next MA's hop does not always ensure the optimal solution. For example, in Figure \ref{GIGMvsCMIP}(a), the output of the LCF algorithm shows that the MA needs to migrate 5 hops after data collection until it reaches the sink. In this context, 5 hops will be at the cost of high energy dissipation and will increase the delay since it carries a large amount of data over many hops. Indeed, it will be more efficient if the nearest source node to the sink becomes the last destination of MA (before reaching the sink). Therefore, in CMIP, the source nodes' visiting order is first determined by the LCF algorithm, then the output itinerary is reversed, such that the last source node in the itinerary will be visited first. 

\begin{figure}[h!]
	\begin{center}
		\includegraphics[width=1\linewidth]{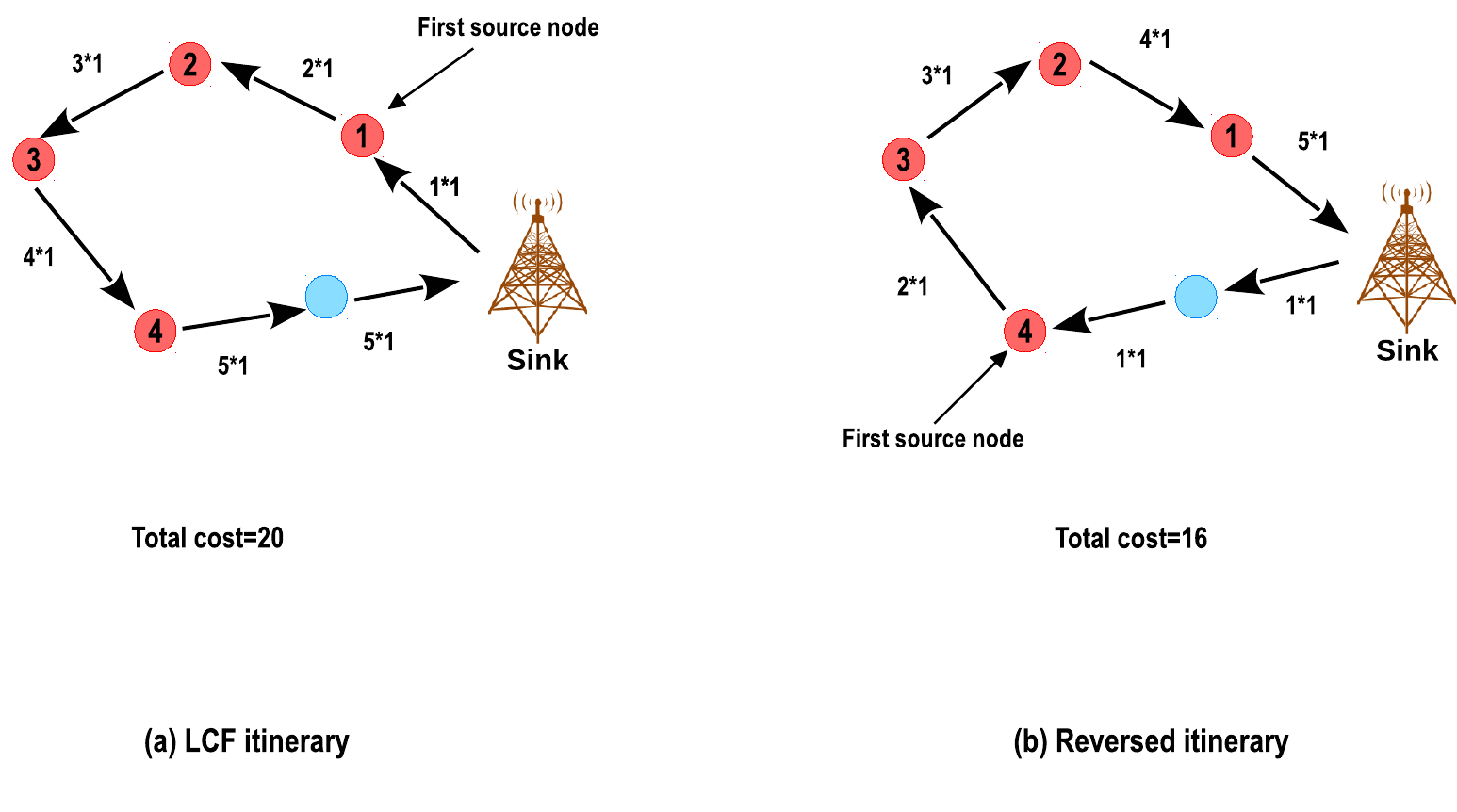}
	\end{center}
	\vspace{-0.5cm}{\textbf{\caption{A comparison of MA itinerary cost: (a) LCF itinerary; (b) Reversed itinerary}\label{reversedLCFitinerary}}}
	\addtocontents{lof}{\protect\addvspace{.5cm}}
\end{figure}

The energy consumption and delay between two source nodes is proportional to the hop count between them with respect to the size of MA. Therefore, the hop count between these two source nodes is the key metric to evaluate the MA's migration cost in terms of energy consumption and delay. In order to evaluate the MA's migration cost in the proposed CMIP, Figure \ref{reversedLCFitinerary} illustrates the comparison of cost between LCF itinerary and reversed itinerary. In this figure, there are four source nodes to be visited by the MA. In Figure \ref{reversedLCFitinerary}(a), the source nodes' visiting order determined by LCF is \textbf{1}, \textbf{2}, \textbf{3} and \textbf{4}, where source node \textbf{1} will be visited first. Let assume that the size of the MA is 1 when dispatched from the sink and it will increase by 1 after each source node visit. By multiplying the current size of MA by the number of hops, the cost of the MA's itinerary can easily be calculated. The results show that the total cost of the LCF itinerary is 20 when source node \textbf{1} is visited first. In contrast, from Figure \ref{reversedLCFitinerary}(b), the total cost of the reversed itinerary is 16. Thus, the reversed itinerary can further decrease the energy consumption and delay. 
       
\begin{algorithm}[h!]
		\caption{Pseudo-code of CMIP approach}
		\label{CMIPalgorithm}
		\textbf{Notation:}\\
		$FSN$ $\leftarrow$ is the farthest source node from sink\\
		$SN_S$ $\leftarrow$ denotes the set of source nodes to be visited by MA\\
		$Num_{iti}$ $\leftarrow$ denotes the number of MA itineraries \\
		$f(FSN, sink)$ $\leftarrow$ is a function that returns the shortest path between the sink and $FSN$ \\
		$f(SN_S, FSN)$ $\leftarrow$ is a function that returns the next source node determined by LCF Algorithm\\
		$FSN_SP$ $\leftarrow$ denotes the MA hop sequence determined by $f(FSN, sink)$ \\
		$VSN_S$ $\leftarrow$ denotes the source nodes' visiting order determined by $f(SN_S, FSN)$ \\
		\textbf{Initialization:}\\
		N $\leftarrow$ Number of sensor nodes\\
		$MA_{DPT}$ $\leftarrow$ The threshold value of the main MA data payload\\
		\textbf{Partitioning the network using k-means algorithm by calculating the distance among N:}\\
		$K$ $\leftarrow$ Number of partitions (specified by user as in GIGM-MIP)\\
		$p$ $\leftarrow$ Set of source nodes in each partition\\
		\For{$j$ = 1 to $K$}{
			$S$ $\leftarrow$ Number of source nodes in set $p$  \\
			$NumMAs$ $\leftarrow$ Number of MAs in set $p$ \\
			\For{$m$ = 1 to $NumMAs$}    {
				Determine the MA itinerary using LCF Algorithm\\	
				Find $FSN$ from the sink using Equation (\ref{FSNlocation})\\
				\uIf{$FSN$ is not the last SN in the MA's itinerary}
				{
					Split the itinerary into two itineraries from $FSN$\\
					
					\For{each splited itinerary}{
						$FirSN$ $\leftarrow$ $FSN$\\
						$FSN_SP$ $\leftarrow$ $f(FSN, sink)$\\
						$VSN_S$ $\leftarrow$ $f(SN_S, FSN)$\\
						add $FSN_SP$ to $VSN_S$ \\	
					}
				}
				
				\Else 
				{
					Reverse the MA itinerary determined by LCF Algorithm \\
				}
			}
		$Num_{iti}$ $\leftarrow$ Reversed and splited MA itineraries 
		}
		
		Return $Num_{iti}$ \\	
\end{algorithm}

In the proposed CMIP, if the FSN is the last source node in the itinerary, the whole itinerary is reversed and there is no cloning process. On the other hand, when the FSN is not the last source node, the reversing procedure will be different. The first part of the itinerary from the sink to the FSN will be reversed where the second part from the FSN to the sink will remain unchanged. Then, each part of the itinerary will be assigned to an individual MA, where all MAs works concurrently. Figure \ref{GIGMvsCMIP}(b) above demonstrate an example of this reversed process. In this figure, the FSN is the source node \textbf{4}, so the itinerary from the sink to the FSN is reversed with a visiting order \textbf{4}, \textbf{3}, \textbf{2}, \textbf{1}, till it reaches the sink. On the other hand, the second part of the itinerary, which begins from \textbf{4} to \textbf{5}, \textbf{6} and \textbf{7}, till it reaches the sink, will remain unchanged. This ensures that the data collection performed by the MA always starts from the farthest source node and proceeds towards the nearest nodes to the sink, which minimizes the delay and energy consumption. Algorithm \ref{CMIPalgorithm} details the pseudo-code of the proposed CMIP approach.

\subsection{Task Duration in CMIP}
As known, the task duration of MIP approach equals the delay of the MA from the dispatching time to it returns to the sink. However, since the CMIP consists of MMA and CMA, the task duration equals the delay from the dispatching time of the MA to the arrival of both MMA and CMA to the sink, whichever is longer. Additionally, since the itinerary of the MMA begins and ends at the sink node and it is also based on Equation (\ref{T_P_MA}), the total delay of its itinerary can be calculated as follows:
\begin{equation}
T_{MMA} = T^{MMA}_{p} + T^{MMA}_{roam} + T^{MMA}_{back}, 
\end{equation}
where $T^{MMA}_{p}$ is the delay of the MMA migration from the sink to the FSN, and $T^{MMA}_{roam}$ is the MMA migration delay from the FSN to the last source node, and $T^{MMA}_{back}$ is the MMA migration delay from the last source node to the sink. As for the case of the CMA, since its itinerary begins at the FSN and ends at the sink, its total task duration can be calculated as follows:
\begin{equation}
\label{T_CMA}
T_{CMA} = T^{MMA}_{p} + T^{CMA}_{roam} + T^{CMA}_{back},
\end{equation}
where $T^{MMA}_{p}$ is the delay incurred due to the MMA's migration from the sink to the FSN plus the delay from FSN to the first source node in CMA's itinerary. $T^{CMA}_{roam}$ is the CMA migration delay from the first source node to the last source node, and $T^{CMA}_{back}$ is the CMA migration delay from the last source node to the sink. Noting that the delay of the $T^{MMA}_{p}$ is added to the delay of CMA in Equation (\ref{T_CMA}) because the delay incurred to reach the FAN is the same for both MMA and CMA.

 \section{Simulation Setup}
The proposed CMIP approach has been implemented and tested on a simulation developed via MATLAB R2017b (student version). The same network model used in \cite{chen2009A,cai2011,aloui2015,wang2015} is adopted which is the most popular network model in data gathering-based MIP. We used the same energy consumption model as in \cite{chen2011,cai2011}. More specifically, a large-scale network, consisting of 800 static nodes dense and uniformly deployed, is considered in the experiments in order to validate the scaling of the proposed CMIP approach. Each node has a transmission range of 60 meters. The sink node has a continuous energy supply and it is located at the center of the network. The generated MAs' itineraries are statically predetermined at the sink node before the MAs are dispatched to the network. In each data gathering task, a random number of source nodes, varying from 10 to 80 by the step of 5, is selected. Each compared MIP algorithm was tested with the same number of selected source nodes. The simulation parameters are listed in Table \ref{table1}.
\begin{table}[h!]
	\caption{Simulation parameters}
	\label{table1}
	\centering
	\begin{tabular}{p{4cm}p{2cm}}
		\hline
		\textbf{Network Parameters}	& \textbf{Value} \\\hline
		Size of network				& 1000 m $\times$ 500 m \\
		Number of deployed nodes	& 800\\
		Number of source nodes		& 10 - 80 \\	
		Transmission range	    	& 60 m\\
		Raw data size	        	& 2048 bits\\ \hline
		\hline
		\textbf{MA Parameters}		& \textbf{Value}\\\hline
		MA processing code 			&	1024 bits\\
		MA accessing delay 			&	10 ms\\
		Raw data reduction ratio	&	0.8\\
		Aggregation ratio			&	0.9\\
		Data processing rate 		&	50 Mbps\\
		\hline
	\end{tabular}
	\label{tab2}
\end{table}
 
\section{Performance Evaluation}
This section evaluates the performance of the proposed CMIP in terms of task duration, Event-to-sink throughput, and energy consumption, where the GIGM-MIP and CL-MIP approaches are chosen for benchmarking. The performance evaluation is done based on two scenarios: (1) a scenario with variable number of source nodes and (2) a scenario with variable aggregation ratio.

\subsection{Impact of Varying Source Nodes}
In this scenario, the number of source nodes is increased from 10 to 80 by a step of 5, while keeping all other parameters, as in Table \ref{table1}, unchanged. In fact, the number of source nodes has a direct impact on the performance of the studied algorithms in terms of energy consumption, Event-to-sink throughput, and task duration because the more the source nodes to be visited, the larger the MA size.

\begin{figure}[h!]
	\begin{center}
		\includegraphics[angle=-90,width=0.9\linewidth]{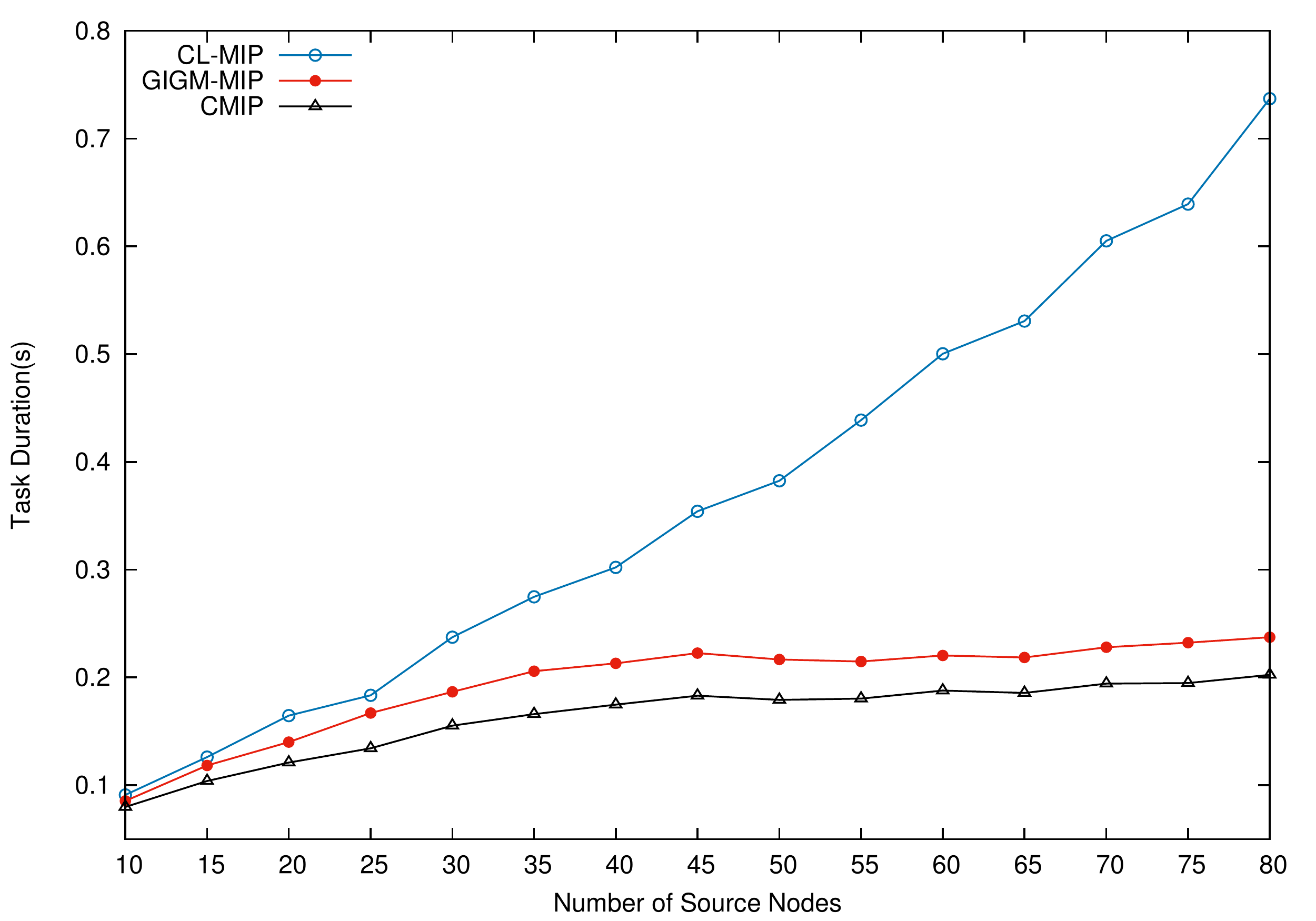}
	\end{center}
	\vspace{-0.5cm}{\textbf{\caption{Task Duration}\label{CMIPTaskDurationch6}}}
	\addtocontents{lof}{\protect\addvspace{.5cm}}
\end{figure}

\begin{figure}[h!]
	\begin{center}
		\includegraphics[angle=-90,width=0.9\linewidth]{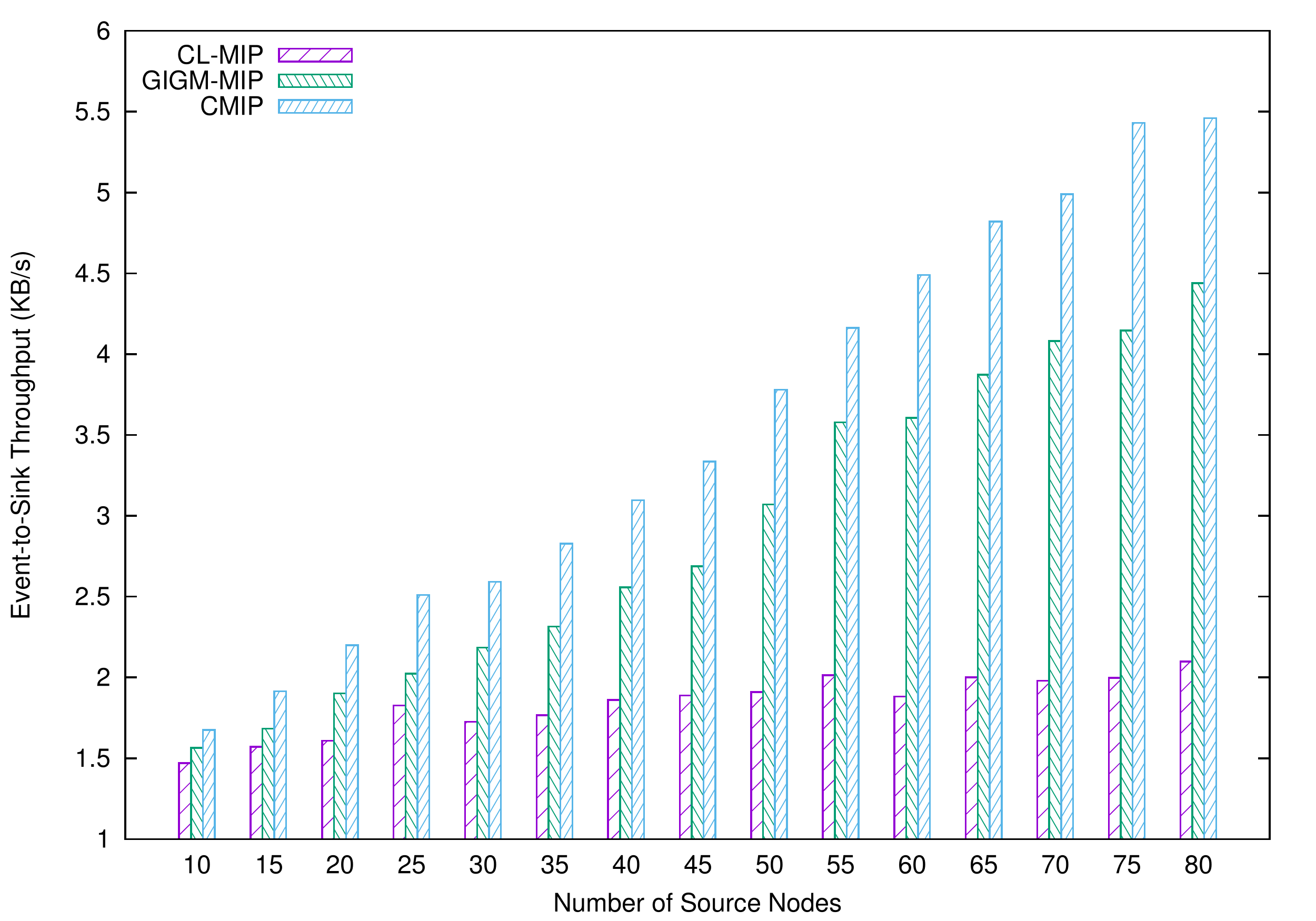}
	\end{center}
	\vspace{-0.5cm}{\textbf{\caption{Event-to-Sink Throughput}\label{CMIPDatacollectedatthesinkch6}}}
	\addtocontents{lof}{\protect\addvspace{.5cm}}
\end{figure}

\begin{figure}[h!]
	\begin{center}
		\includegraphics[angle=-90,width=0.9\linewidth]{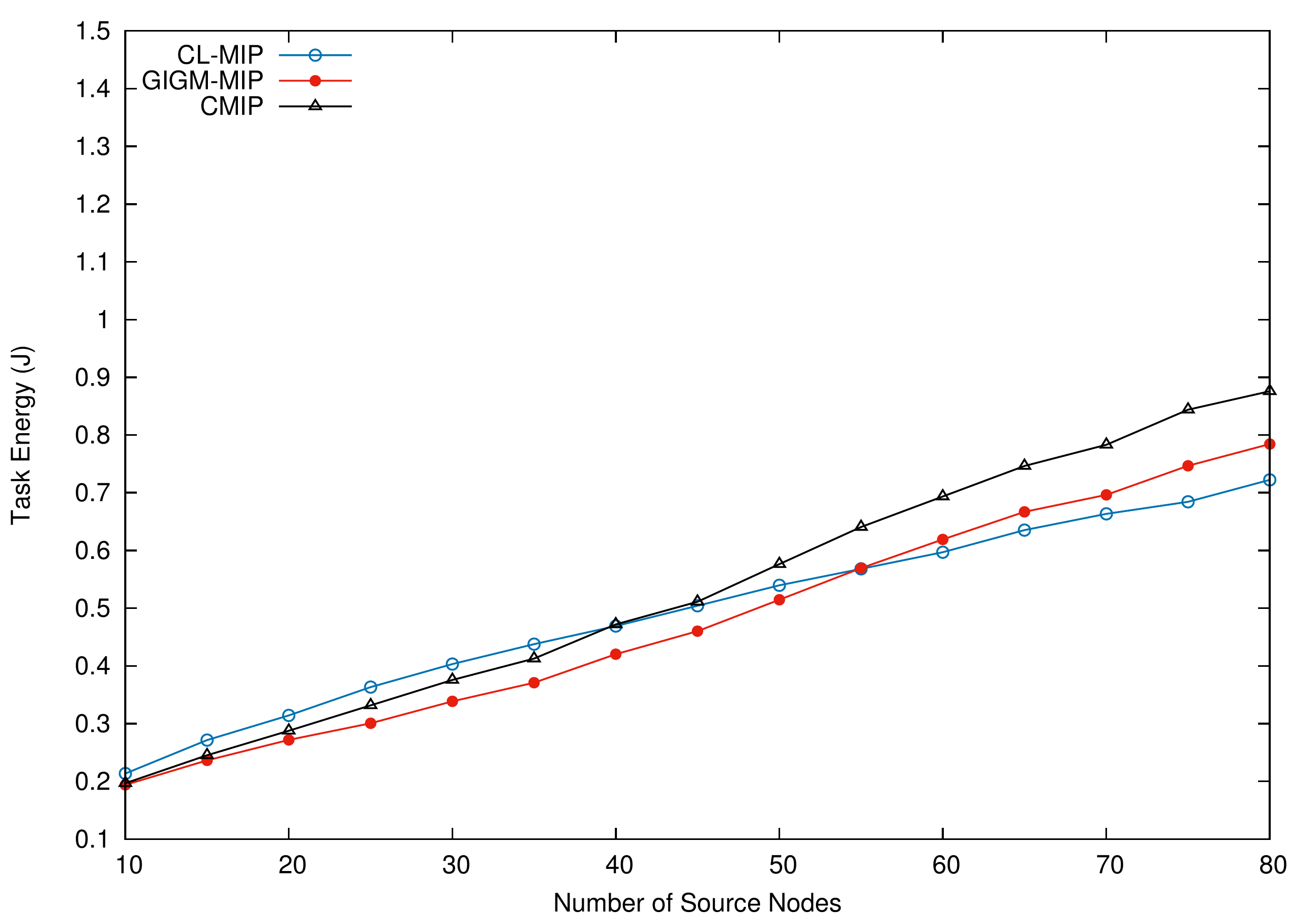}
	\end{center}
	\vspace{-0.5cm}{\textbf{\caption{Energy consumption}\label{CMIPEnergych6}}}
	\addtocontents{lof}{\protect\addvspace{.5cm}}
\end{figure}

Figures \ref{CMIPTaskDurationch6}, \ref{CMIPDatacollectedatthesinkch6}, and \ref{CMIPEnergych6} show the impact of varying the number of source nodes on the task duration, Event-to-sink throughput and energy consumption, respectively. In Figure \ref{CMIPTaskDurationch6}, it is clear that the proposed CMIP achieves the minimum task duration by about 56\% and 16\% reduction compared to CL-MIP and GIGM-MIP approaches, respectively. Moreover, CMIP decreases the task duration linearly compared to GIGM-MIP when the number of source nodes increases, as a result of using the cloning mechanism, which splits the MA itineraries that have a large number of source nodes. This split operation helps each distributed MA to take less number of source nodes, which in turn helps them to complete the data gathering journey much faster.

Moreover, the reduction in task duration noticed in CMIP also improves the Event-to-sink throughput. In WSNs, the Event-to-sink throughput $E$ is defined as the number of packets received at the sink over the time period $T$ \cite{wang2007improving}. Since the MAs in MA-based data gathering algorithms are responsible for collecting and delivering the aggregated data to the sink within the task duration, thus $E$ can be calculated as follows:
\begin{equation}
	E = \frac{T_{Data}}{T_{Duration}},
\end{equation}
where $T_{Data}$ is the total aggregated data that has been successfully delivered to the sink by all MAs and $T_{Duration}$ is the task duration from beginning till the end.

Figure \ref{CMIPDatacollectedatthesinkch6} shows that the CMIP is able to outperform the compared schemes (GIGM-MIP and CL-MIP) by attaining higher Event-to-sink throughput due to the ability of the involved MAs in CIMP to complete their round-trip of delivering the data to the sink with lower task duration. In contrast, GIGM-MIP and Cl-MIP show lower Event-to-sink throughput due to the longer time needed for the MAs to accomplish their data gathering process. In GIGM-MIP and Cl-MIP, some MAs are assigned to a large number of source nodes, which in turn increases the number of hops and MA data payload size that results in task duration increase.

Figure \ref{CMIPEnergych6} presents the energy consumption needed to accomplish the task of CMIP compared to GIGM-MIP and CL-MIP. In MIP, the task energy consumption is the accumulated energy spent by all distributed MAs to perform the data gathering task from all source nodes \cite{cai2011,gavalasEnergy}. It includes the energy spent on transmitting, receiving, and exchanging control message. Thus, the total task energy consumption can be calculated as follows:
\begin{equation}
	C_{total} =\sum_{t=1}^{|I|} IC^t,
\end{equation}
where $IC^t$ is the energy cost of itinerary $I^t$ covered by the MA, and $IC^t$ can be simplified to: 
\begin{equation}
	IC^t =\sum_{j=1}^{|I^t|} (jdf+pc)c_{i,j},
\end{equation}
where $|I^t|$ represents the number of visited nodes in the itinerary $I^t$ by the relevant MA, $j$ is the visited sensor node, $jdf$ is the size of data collected by the MA at the sensor node $j$ after it is aggregated by a ratio of $f$,  $pc$ is the MA's initial size (processing code plus MA packet header), and $c_{i,j}$ is the energy consumption of the MA to migrate from sensor node $i$ to sensor node $j$. Noting that $j$ could act as a source node (has data to be collected by the MA) or as an intermediate node (forwarding node).

The proposed CMIP has a compatible performance with GIGM-MIP when there are 10-20 source nodes. However, it has lower energy consumption compared to CL-MIP scheme. As the number of source nodes increase, CMIP starts to consume higher energy compared to both GIGM-MIP and CL-MIP schemes because CMIP starts to clone its MA for the itineraries that have a larger number of source nodes which in turn results in a slightly higher number of hops. It is worth noting that the increase in hops does not affect the MA round trip since the data collected by the MMA and CMA is well distributed. Consequently, the MAs could successfully complete their tasks with less delay and high Event-to-sink throughput.

\subsection{Impact of Varying Aggregation Ratio:}
This section evaluates the impact of varying data aggregation ratio $f$ on the performance of CMIP. In this scenario, the varied $f$ ratio represents different redundancy and compression function of the collected data. Thus, different sizes of collected data can be delivered to the sink, which has a direct impact on the proposed CMIP in terms of task duration, Event-to-sink throughput, and energy consumption. In this experiment, $f$ ratio is varied from 0.1 to 0.9 for 80 source nodes.

\begin{figure}[h!]
	\begin{center}
		\includegraphics[angle=-90,width=0.9\linewidth]{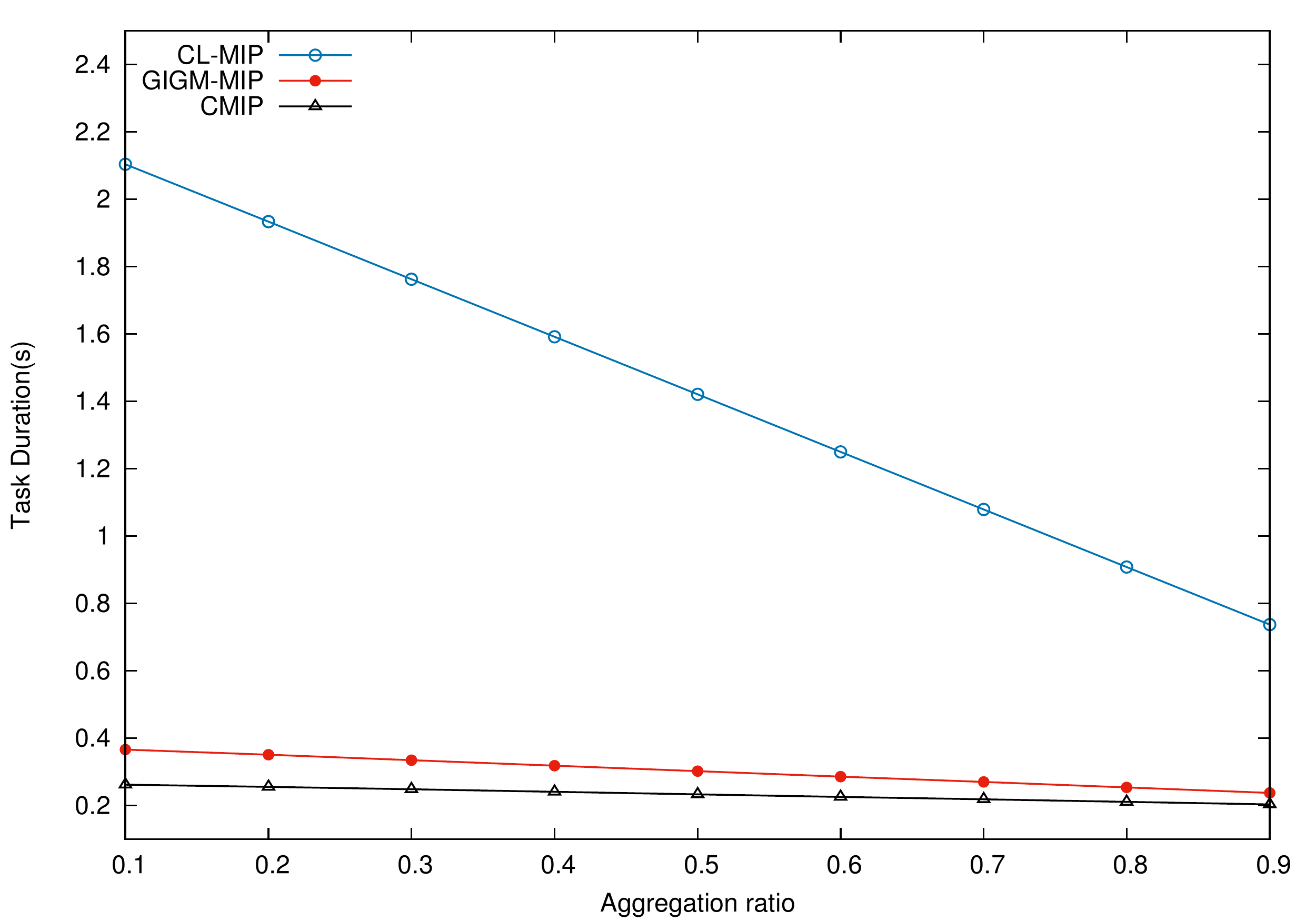}
	\end{center}
	\vspace{-0.5cm}{\textbf{\caption{Task Duration}\label{CMIPTaskDurationvariedaggch6}}}
	\addtocontents{lof}{\protect\addvspace{.5cm}}
\end{figure}

\begin{figure}[h!]
	\begin{center}
		\includegraphics[angle=-90,width=0.9\linewidth]{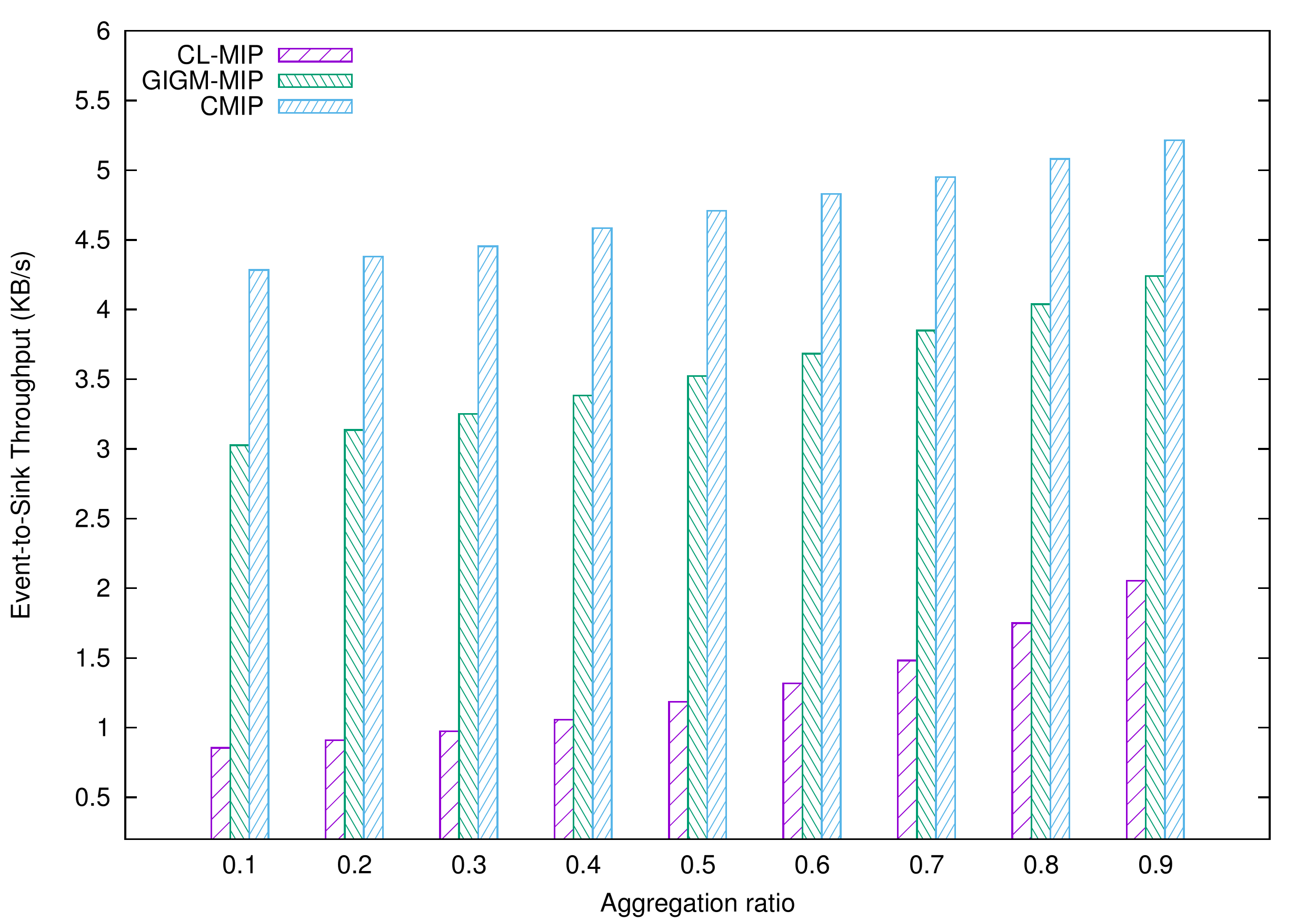}
	\end{center}
	\vspace{-0.5cm}{\textbf{\caption{Event-to-Sink Throughput}\label{CMIPDatacollectedatthesinkvariedaggch6}}}
	\addtocontents{lof}{\protect\addvspace{.5cm}}
\end{figure}

\begin{figure}[h!]
	\begin{center}
		\includegraphics[angle=-90,width=0.9\linewidth]{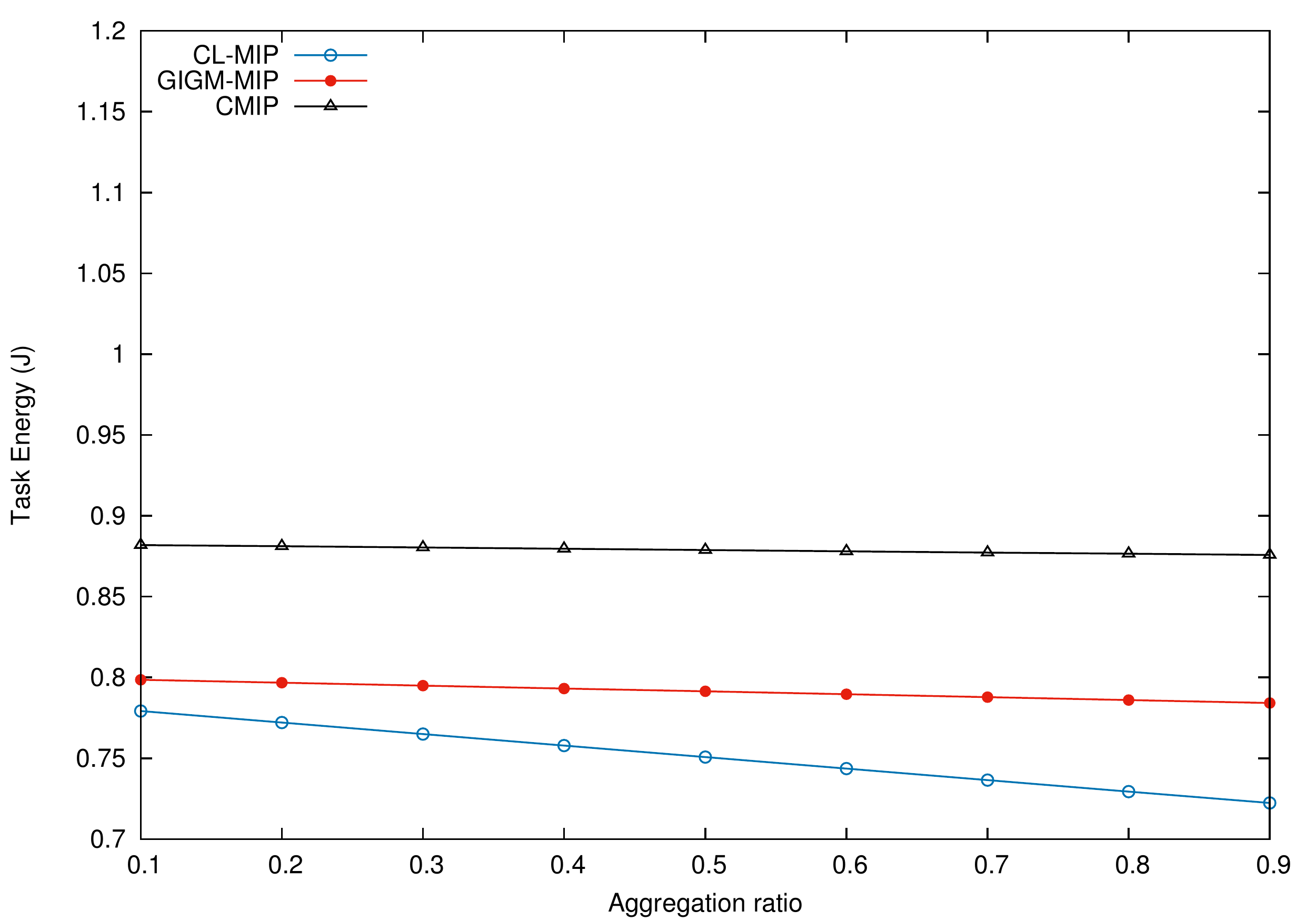}
	\end{center}
	\vspace{-0.5cm}{\textbf{\caption{Energy consumption}\label{CMIPEnergyvariedaggch6}}}
	\addtocontents{lof}{\protect\addvspace{.5cm}}
\end{figure}

Figures \ref{CMIPTaskDurationvariedaggch6}, \ref{CMIPDatacollectedatthesinkvariedaggch6} and \ref{CMIPEnergyvariedaggch6} show the impact of varying the ratio of $f$ on task duration, Event-to-sink throughput and energy consumption, respectively. As shown in Figure \ref{CMIPTaskDurationvariedaggch6}, the proposed CMIP outperforms GIGM-MIP and CL-MIP in terms of task duration, although the size of MA becomes larger whenever the value of $f$ becomes smaller. This is due to the cloning strategy deployed by the CMIP, which facilitates the collaboration between the involved MAs (MMA and CMA) allowing them to collect larger data collaboratively in a shorter duration of time.

On the other hand, Figure \ref{CMIPDatacollectedatthesinkvariedaggch6} shows that the Event-to-sink throughput of CMIP is higher compared to GIGM-MIP and CL-MIP schemes. Although the size of the collected data changes on varying $f$, CMIP still performs better than the other schemes due to the large amount of data collected by the MMA and CMA within a shorter time, which improves the Event-to-sink throughput. As the size of the collected data in GIGM-MIP and CL-MIP increases, the MA size increases. Thus, receiving and transmitting an MA with large data at any hop consumes a significant amount of time, which increases the total task duration and directly decreases the Event-to-sink throughput.

Additionally, receiving and transmitting an MA that carries a large amount of data at any hop consumes more energy, which may shut down that hope due to having a low level of energy. As a result, the MA with its carried data will be lost, which negatively affects the Event-to-sink throughput. However, the increase in energy consumption caused by the large size of MA has a negligible impact on the Event-to-sink throughput in CMIP, as shown in Figure \ref{CMIPEnergyvariedaggch6}, due to the use of the cloning strategy that deploys two MAs to gather the data from a large number of source nodes. Thereby, the collected data will be divided over the distributed MAs allowing them to complete their task with minimum time and high Event-to-sink throughput.

\section{Conclusion}
In this paper, a Clone Mobile-agent Itinerary Planning (CMIP) approach has been proposed. The proposed CMIP adopt the cloning concept of MA to reduce the task duration when the MA's itinerary has a large number of source nodes to be visited. The CMIP mitigates this issue by splitting the itinerary into sub-itineraries, each of which is assigned to an individual MA. Moreover, a reversed MA's itinerary is proposed to further decrease the energy consumption and task duration. Further, simulation experiments have been conducted to evaluate the performance of CMIP. The results show that CMIP significantly outperforms the compared schemes in terms of task duration and Event-to-sink throughput.

\begin{IEEEbiography}[{\includegraphics[width=1in,height=1.25in,clip,keepaspectratio]{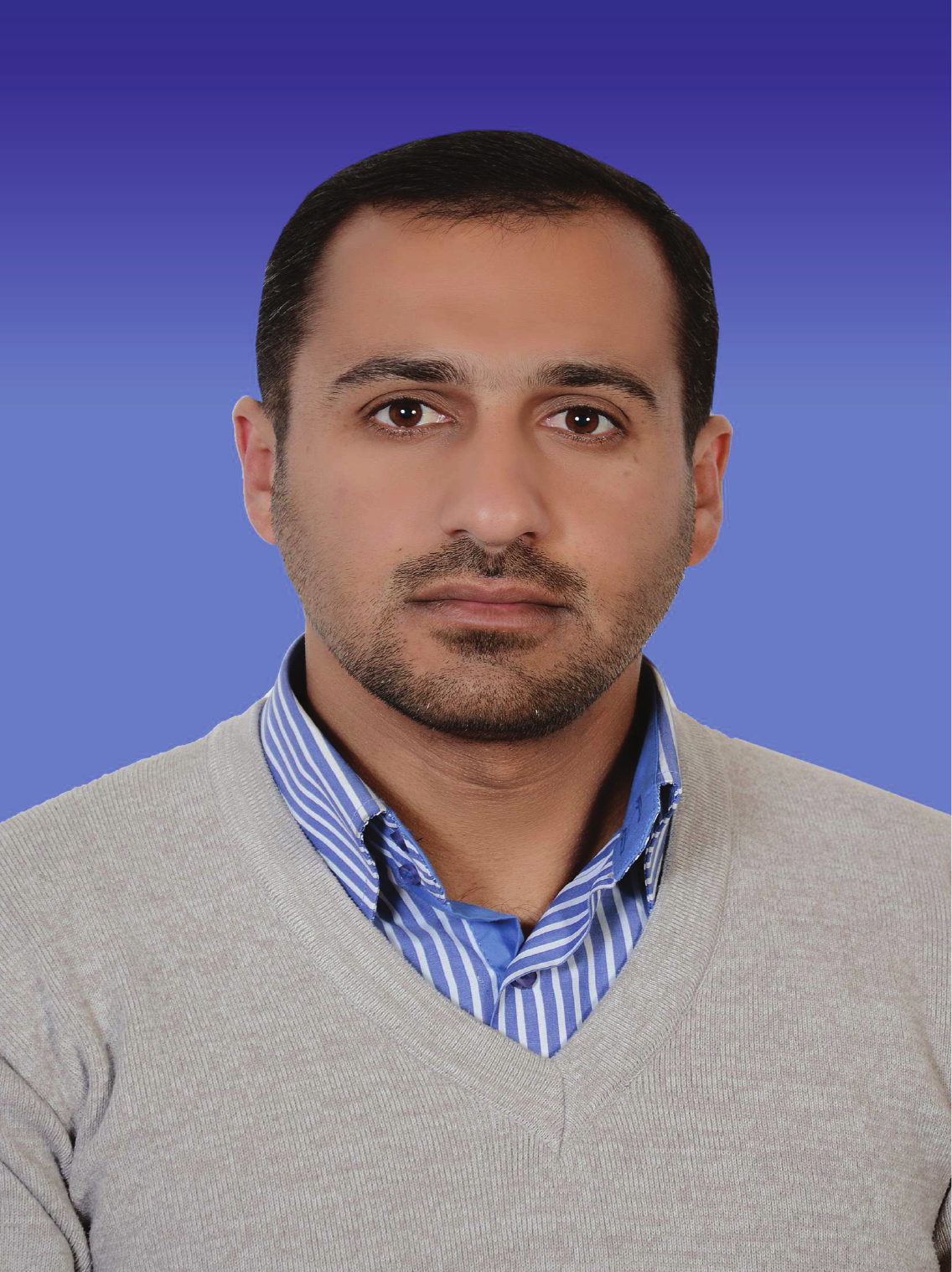}}]{Huthiafa Q Qadori}
received his bachelor degree in computer science from Faculty of Computer Science and Information Technology, Universiti of Al-Anbar, Iraq, in 2006, and his master of Information Technology from University Tenaga National (Uniten), Malaysia, in 2012. He is currently pursuing his Ph.D. degree at the Department of Communication Technology and Networks, Universiti Putra Malaysia. His research interests are in computer networks, wireless sensor networks, routing algorithms and the Internet of things.
\end{IEEEbiography}

\begin{IEEEbiography}[{\includegraphics[width=1in,height=1.25in,clip,keepaspectratio]{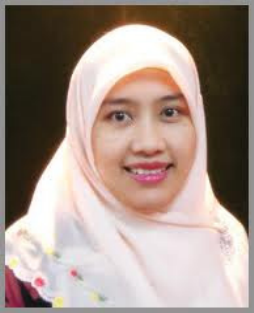}}]{ZURIATI AHMAD ZUKARNAIN}
(M'10) received her bachelor and master degrees in physics and education from Universiti Putra Malaysia (UPM) in 1997 and 2000, respectively, and her Ph.D. degree in quantum computing and communication from the University of Bradford, U.K., in 2005. Now, she is a professor in the Faculty of Computer Science and Information Technology, UPM. She was appointed as Head of Department for Communication Technology and Networks from 2006 to 2011. She also being appointed as the Head of the Section of High-Performance Computing, Institute of Mathematical Research, UPM, from 2012 to 2015. She taught several courses for the undergraduate students such as data communication and networks, distributed system, mobile and wireless networks, network security, computer architecture, and assembly language. For postgraduate students, she taught few courses such as advanced distributed computing and research method. Also, she is a member of the IEEE. Her areas of interest are computer networks, distributed system, mobile and wireless networks, network security, quantum computing, and quantum cryptography.
\end{IEEEbiography}

\begin{IEEEbiography}[{\includegraphics[width=1in,height=1.25in,clip,keepaspectratio]{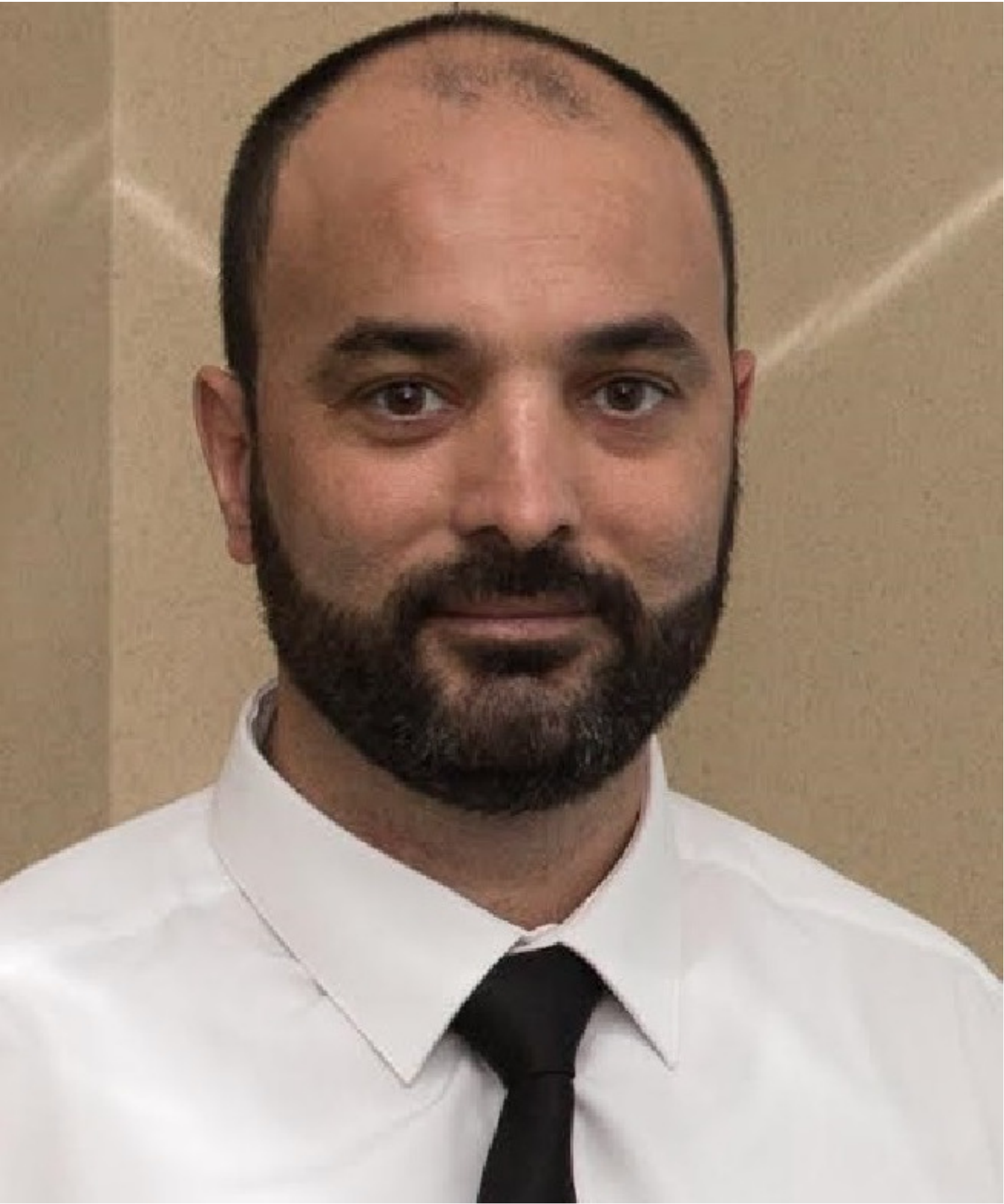}}]{Mohamed A. Alrshah}
	(M'13--SM'17) received his BSc degree in Computer Science from Naser University - Libya, in 2000, and his MSc and Ph.D. degrees in communication technology and networks from Universiti Putra Malaysia in May 2009 and Feb 2017, respectively. Now, he is a Senior Lecturer in the Department of Communication Technology and Networks, Faculty of Computer Science and Information Technology, Universiti Putra Malaysia (UPM). Also, he is a senior member of the IEEE. He has published a number of articles in high impact factor scientific journals. His research interests are in the field of high-speed TCP protocols, high-speed wired and wireless network, parallel and distributed algorithms, WSN, IoT, and cloud computing.
\end{IEEEbiography}

\begin{IEEEbiography}[{\includegraphics[width=1in,height=1.25in,clip,keepaspectratio]{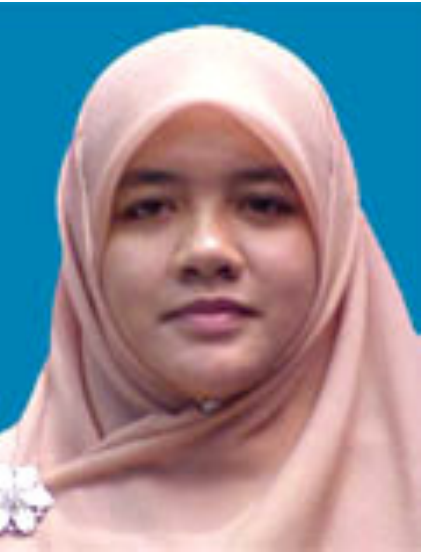}}]{Zurina Mohd Hanapi}
(M'10) received her bachelor in Computer and Electronic System Engineering from Strathclyde University in 1999, and her master in Computer and Communication Systems Engineering from UPM in 2004, and her Ph.D. in Electrical, Electronic, and System Engineering from Universiti Kebangsaan Malaysia in 2011. Now, she is an Associate Professor in the Faculty of Computer Science and Information Technology, University Putra Malaysia (UPM). She has received an Excellence Teaching Awards in 2005, 2006 and 2012 and she has received a silver medal in 2004 and bronze medal in 2012. She is a leader of some research projects and she has published many conference and journal papers and she is a member of the IEEE. Her research interests in Routing, Wireless Sensor Network, Wireless Communication, Distributed Computing, Network Security, Cryptography, and Intelligent Systems.
\end{IEEEbiography}
\newpage
\begin{IEEEbiography}[{\includegraphics[width=1in,height=1.25in,clip,keepaspectratio]{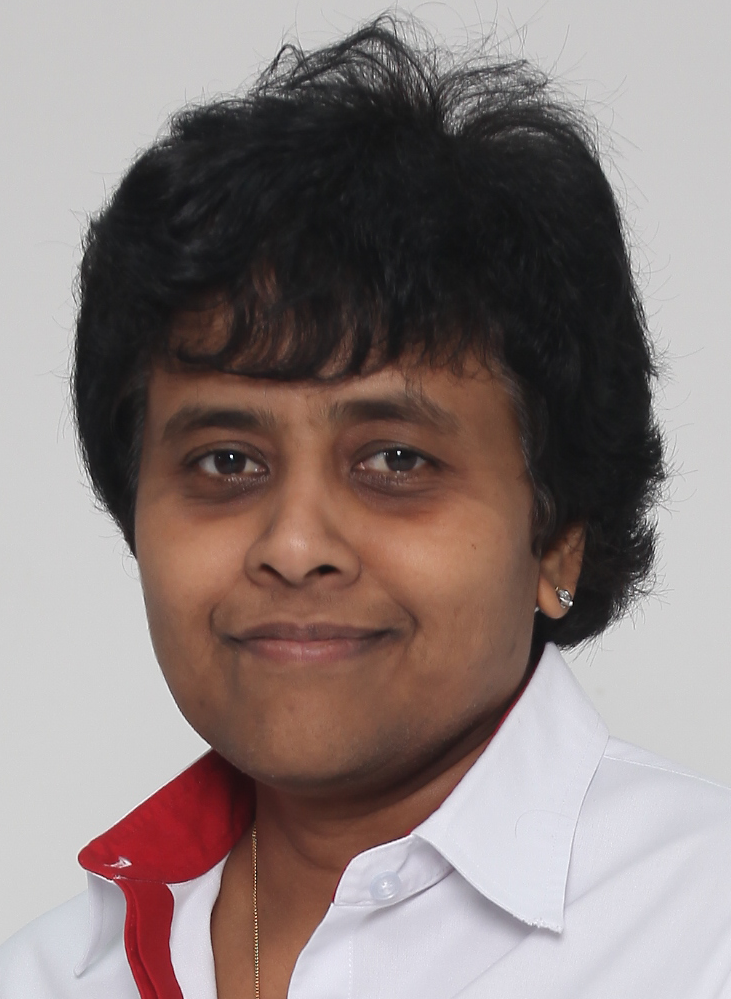}}]{Shamala Subramaniam}
(M'10) received her Bachelor, Master, and Ph.D in Computer Science from University Putra Malaysia (UPM) in 1996, 1999, and 2002, respectively. Now, she is a professor in the Department of Communication Technology and Network, Faculty of Computer Science and Information Technology, Universiti Putra Malaysia. Also, she is a member of the IEEE. Her research interests are Computer Networks, Simulation and Modeling, Scheduling and Real-Time System.
\end{IEEEbiography}
\EOD

\end{document}